\documentclass{svjour3}
\usepackage{hyperref}
\usepackage{latexsym}
\usepackage{graphicx}
\usepackage{mathrsfs}
\usepackage{amssymb,amsmath,bm}
\usepackage{amsfonts}
\usepackage[normalem]{ulem}

\usepackage[mathlines,displaymath]{lineno}
\begin{document}
\title{On the existence of quantum signature for quantum messages}
\author{Qin Li         \and        Wai Hong Chan   \and        Chunhui Wu  \and    Zhonghua Wen}


\institute{Qin Li\at
              MOE Key Laboratory of Intelligent Computing and Information Processing, College of Information Engineering, Xiangtan University, Xiangtan 411105, China
              \at
            %  Fax: +123-45-678910\\
%               \email{liqin805@163.com}           %  \\
\and
  Wai Hong Chan \at
  Department of Mathematics and Information Technology, The Hong Kong Institute of Education, Hong Kong
\and Chunhui Wu \at
Department of Computer Science, Guangdong University of Finance, Guangzhou 510521, China
\and Zhonghua Wen \at
 College of Information Engineering, Xiangtan University, Xiangtan 411105, China
}

\date{Received: / Revised version: }

\maketitle

\begin{abstract}
Quantum signature (QS) is used to authenticate the identity of the originator, ensure data integrity and provide non-repudiation service with unconditional security. Depending on whether a trusted third party named arbitrator is involved or not, QS is classified as arbitrated QS and true QS. This paper
studies existence problem about the two kinds of QS and contributes to two points: 1) a basic framework is provided
to analyze the possibility of arbitrated QS on signing quantum messages; 2) disagreement between the impossibility of true QS and an existing true QS scheme
is solved.
\end{abstract}

\section{Introduction}

Digital signature,  as an analogy to hand-written signature for authenticating the origin of a message and ensuring the message not modified during transmission, is an essential cryptographic primitive. It has been being widely used in various security-related fields, particularly in secure electronic commerce.
As Rivest, one of the three inventors of the RSA algorithm, predicted, digital signature may become one of the most fundamental and useful inventions of modern cryptography \cite{Rivest90:DS}. However, all of the existing classical (digital) signature schemes were threatened by last-increasing power of computers and innovative techniques such as quantum computation since their security depends on the difficulty of solving some hard mathematical problems. Once quantum computers is successfully built, most of classical signature schemes will be cracked through Shor's algorithm \cite{Shor:AQC:FOCS94}. On the other hand, development of quantum physics has thrown some light on the study of cryptography for obtaining unconditional security \cite{BB:QKD:CSSP84,Ekert:QKD:PRL91,Brassard96:QS,LC:QKD:Science99,Bennett00:QS,Mayers01:QS}. So, researchers turn to investigate quantum counterpart of classical signature. They hope that quantum signature (QS) can provide unconditional security, namely the attacker (or the malicious receiver) cannot forge the signature, and the signatory cannot deny the signature even though unlimited computing resources are available.

QS is expected to sign both classical and quantum messages, and the form of each quantum message can be a known quantum state or an unknown one. Since known quantum states can be characterized with classical information, only the quantum messages of the form of unknown quantum states are considered in this paper. In 2001, Gottesman and Chuang proposed the first QS scheme based on quantum one-way function, which is unconditionally secure even against quantum attacks \cite{Gottesman:QS:arXiv01}. But this scheme works only on classical messages, and seems not practicable as it use up $O(m)$ qubits of the public key for signing an $m$-bit message. Subsequently, Barnum \emph{et al}. showed that unconditionally secure QS for quantum messages cannot be achieved \cite{Barnum:AQM:FOCS02}. This negative result did depress some researchers pursuing positive application of QS. Fortunately, Zeng and Keitel presented a QS scheme signing both classical and quantum messages by introducing a trust third party named arbitrator \cite{Zeng:AQS:PRA02} (this kind of QS was named arbitrated QS thereafter), and the scheme was improved later \cite{Curty:AQS:PRA08,Zeng:AQS:PRA08}. Afterwards, Li \emph{et al}. observed that the GHZ states used in \cite{Zeng:AQS:PRA08} could be replaced by Bell states, and then put forward a more efficient scheme in \cite{Qin:Sign:PRA09}. But, Zou \emph{et al}. proved the two schemes proposed in \cite{Zeng:AQS:PRA08} and \cite{Qin:Sign:PRA09} are both insecure because they could be repudiated by the receiver Bob, and further presented two arbitrated QS schemes to fix the defect \cite{Zou:Sign:PRA10}. Some other typical arbitrated QS schemes were also presented since the study of arbitrated QS initiated by Zeng and Keitel \cite{Lee:Sign:PLA04,Lu:AQS:ISCIS04,Lu:QDS:ICACT05,Wang:AQS:ICACT06,Yang:AQS:OC10}.

Recently, almost all typical arbitrated QS schemes designed for quantum messages were found to have security problems of different extents \cite{Gao:Sign:PRA11,Choi:Sign:pra11,Sun:Sign:upd11,Hwang:Sign:pra12,CLH:AQS:OC11,QCL:AQS:JPA13}. So, weather QS can provide unconditional security for quantum messages is the key problem many researchers cared. To solve this doubt, we shall present a detailed analysis of the reasons why unconditionally secure arbitrated QS for quantum messages is possible and this does not contradict with Barnum \emph{et al}.'s conclusion. Note that some related preliminary reasons were simply mentioned by Li \emph{et al}. in \cite{Qin:Sign:PRA09}. In addition, Zeng \emph{et al}. presented a true QS scheme in 2007 and claimed it can sign quantum messages with unconditional security \cite{Zeng:CVQS:IJQI07}. This result stimulated the nerves of researchers in the field, and a question emerge: is Barnum \emph{et al}.'s conclusion in \cite{Barnum:AQM:FOCS02} wrong or Zeng \emph{et al}.'s scheme in \cite{Zeng:CVQS:IJQI07} insecure? We shall answer this question by showing the insecurity of Zeng \emph{et al}.'s scheme.

The rest of the paper is organized as follows. Section~\ref{sec:1} briefly reviews Burnum \emph{et al}.'s conclusion about the impossibility for signing quantum messages. Then the arbitrated QS is shown to be able to sign quantum messages in Sec.~\ref{sec:2}, and the disagreements between the impossibility of true QS and an existing true QS scheme for quantum messages is solved in Sec.~\ref{sec:3}. The last section concludes the paper.

\section{Review of a negative result on signing quantum messages}
\label{sec:1}

This section briefly reviews Barnum \emph{et al}.'s conclusion claiming that signing quantum messages is impossible to realize \cite{Barnum:AQM:FOCS02}.

In \cite{Barnum:AQM:FOCS02}, Barnum \emph{et al}. proved a theorem that quantum authentication implies encryption. This means that any scheme ensuring the authenticity of quantum messages must also encrypt them almost perfectly. However, in a QS scheme, the receiver should learn something about the contents of the quantum message but is not allowed to changed it. It follows that the theorem results in the impossibility of signing quantum messages: any non-trivial information obtained from encrypted quantum messages is only possible at the cost of introducing disturbance to them which implies the authenticity of quantum messages is destroyed.

To be better understanding, one can assume the receiver is allowed to efficiently extract the original quantum message $\rho$, then it is easy to show the receiver can generate a valid signature of a new message $\rho'$ favorable to him by the following steps. First suppose the receiver can extract the original message $\rho$ via the transformation $U$ and leave the auxiliary state as $\varphi$ which may not be hold entirely by the receiver. Since $\rho$ should have been entangled with a reference system, $\varphi$ must be independent of $\rho$. Then the receiver implements the transformation $U^\dagger$, the inverse process of $U$, on $\rho'$ and his part of $\varphi$ to get a valid signature. Obviously, this contradicts with the security of QS.

\section{Possibility of arbitrated QS for quantum messages}
\label{sec:2}

In this section, we analyze why it is possible to design an arbitrated QS scheme to sign quantum messages. It does not disagree with Barnum \emph{et al.}'s conclusion \cite{Barnum:AQM:FOCS02} and can provide unconditional security if it is properly devised. We also give a basic framework of such a scheme as an example.

As mentioned in Sec.~\ref{sec:1}, Barnum \emph{et al.} showed any protocol allowing a receiver to read a quantum message also allows the receiver to modify the message without the risk of being detected, therefore all potential receivers of an authenticated message must be trustworthy. Obviously this will not happen for a general QS protocol. However, in an arbitrated QS scheme, the arbitrator is always supposed to be trusted by both signatory and receiver. We assume that the real recipient of the authenticated message is the arbitrator who is in charge of the verification of the signature. After verifying the signature, the arbitrator can send a parameter to indicate whether the signature is valid. The receiver would obtain the indication parameter, and only check whether the parameter and other information come from the real arbitrator.

Recently, almost all existing arbitrated QS schemes were cracked  \cite{Gao:Sign:PRA11,Choi:Sign:pra11,Sun:Sign:upd11,Hwang:Sign:pra12,CLH:AQS:OC11,QCL:AQS:JPA13}, but it does not mean that arbitrated QS cannot provide unconditional security. The failures of the previous schemes are mainly due to imperfections of design. For example, some schemes just employed quantum one-time pad to encrypt, but ignored to authenticate the transmitted quantum messages. Quantum encryption does not imply authentication, even though the converse is true \cite{Barnum:AQM:FOCS02}.

Based on the above analysis, we give a basic framework of an unconditional secure arbitrated QS scheme for quantum messages, which
is composed of the following three phases.
\begin{itemize}
\item {\it The initial phase:} At first, the signatory Alice shares a key $K_{A}$ with the arbitrator using an unconditionally secure quantum key distribution protocol. The receiver Bob also has a key $K_{B}$ shared with the arbitrator in the same way.

\item {\it The signing phase:} Alice generates the signature $Sig_{K_A}(P)$ of the message $P$ via a noncommutative transformation which means that there does not exist a unitary operation $U$ to satisfy $U(Sig_{K_A}(P))=Sig_{K_A}(UP)$. Then Alice computes $\sigma=Aut(Sig_{K_A}(P), P)$ for authentication. Note that $Aut(\cdot)$ denotes that unconditionally secure authentication is used such as the quantum authentication scheme given in \cite{Barnum:AQM:FOCS02} for quantum information, and Wegman-Carter authentication scheme for classical information in \cite{MS:MAC:81}. The authenticated signature state $\sigma$ is sent to Bob.

\item {\it The verification phase:} After Bob receives $\sigma$, he can implement the verification with the aid of the arbitrator using the following three steps:
\begin{itemize}
\item[(1)] In the beginning, Bob makes a random transformation on $\sigma$ to obtain $Trans_{K_B}(\sigma)$ according to the key $K_B$. Then he produces the authenticated state $Y=Aut(Trans_{K_B}(\sigma))$ and transmits it to the arbitrator.
\item[(2)] After the arbitrator received $Y$, she checks the authenticity of $Trans_{K_B}(\sigma)$. If there is anything wrong, the arbitrator would abort the protocol immediately; otherwise, the arbitrator would recover $\sigma$ from $Trans_{K_B}(\sigma)$ using the key $K_B$ shared with Bob. Similarly, the arbitrator examines whether $Sig_{K_{A}}(P)$ and $P$ are tamped or not. If not, the arbitrator would verify whether $Sig_{K_A}(P)$ is a valid signature of the message $P$. If the verification process is passed, the arbitrator sets the verification parameter $r=1$; otherwise, sets $r=0$. Finally, the arbitrator computes $T=Aut(P,Sig_{K_A}(P), r)$ and sends it to Bob.
\item[(3)] Bob authenticates what he received. If the authentication test is passed and $r=1$, he would accept $Sig_{K_A}(P)$ as the signature of $P$.
\end{itemize}
\end{itemize}
The arbitrated QS scheme of the above framework is secure in terms of two aspects: impossibility of forgery and non-repudiation of signatory. Without the knowledge about signing key, an attacker cannot directly produce a signature of a new quantum message directly. Furthermore, as the signature process is noncommutative, an attacker cannot change the received signature state to be that of another quantum message favorable for him by implementing appropriate unitary operations. The key information of the signatory is included in a signature state, which cannot be changed due to the use of authentication during the verification phase, so the signatory cannot deny what she has signed. In addition, under the assumption that the arbitrator is trustworthy and is the real person who can verify $Sig_{K_A}(P)$ with all the information he holds, thus the arbitrated scheme under the proposed framework does not disobey Barnum \emph{et al.}'s conclusion. This also tells that the verification made by the arbitrator is indispensable to an arbitrated QS scheme.

\section{Building a bridge between two contradictory results on signing quantum messages}
\label{sec:3}

From the above section, we know Barnum et al.'s conclusion \cite{Barnum:AQM:FOCS02} about the impossibility of signing quantum messages does not include arbitrated QS, so it just aimed at true QS. However, Zeng et al. presented a true QS scheme recently in \cite{Zeng:CVQS:IJQI07}, which was claimed to be able to sign quantum messages with unconditional security. Then people wondered whether Barnum et al.'s conclusion is wrong or Zeng et al.¡¯s scheme is not secure. In this section, we clarify this question. At first, we review Zeng \emph{et al}.'s true QS scheme \cite{Zeng:CVQS:IJQI07}. Then, we show the insecurity of the scheme by attacking it successfully using similar method presented by Barnum \emph{et al}. in \cite{Barnum:AQM:FOCS02}.

\subsection{Review of the true QS scheme proposed by Zeng et al.}

Zeng \emph{et al}. proposed a true QS scheme for the purpose of signing quantum messages based on a suitable one-way function recently \cite{Zeng:CVQS:IJQI07} and claimed that the scheme is unconditionally secure. We briefly describe the three main phases of their scheme as follows (More details can be found in \cite{Zeng:CVQS:IJQI07}).

\begin{itemize}
\item In the initial phase, the main goal is to generate a signing key $K_s$ and a verification key $K_v$ by constructing a one-way function $G:\{L,X,T_{ij}\}\rightarrow\{U,\|T\|^{1/2}\}$, where $L$ is a linear transformation mapping $x=(x_0,x_1,...,x_{k-1})\in \mathbb{R}^k$ to $y(X)=y_0(x)=[x_0,y_1(x),...,y_{2k-1}(x)]^{T}\in\mathbb{R}^{2k}$ and making any $k$-element subset of $\{x_0,y_1,...,y_{2k-1}\}$ linearly independent, $T$ makes $T[y_{r_1},y_{r_2},...,y_{r_k}]^T=T[x_0,y_{r_{k+1}},...,y_{r_{2k-1}}]^T$, and $U$ satisfies $U|y_{r_1}\rangle_{r_1} |y_{r_2}\rangle_{r_2}...|y_{r_k}\rangle_{r_k}=|x_0\rangle_{r_1}|y_{r_{k+1}}\rangle_{r_2}$ $\ldots|y_{r_{2k-1}}\rangle_{r_k}$. $K_s$ is expressed as $K_s=\{L,X\}$ and $K_v$ is set as $K_v=\{U,\|T\|^{1/2}\}$.
\item In the signing phase, according to $K_s$, the signatory Alice prepares $2k-1$ ancilla states $|\omega(X)\rangle=|y_1(X)\rangle_1...|y_{2k-1}(X)\rangle_{2k-1}$ and encodes the message state $P$ with a wave function $\langle x_0|P\rangle$ as $|\tilde{S}\rangle=\int |P\rangle|\omega\rangle dX$. Then Alice prepares a two-particle entangled state $|\tilde{\Omega}\rangle=\int_\mathbb{R}|y_{k+1}\rangle_{r_2}|y_{k+1}\rangle_{r_{k+1}}dx$ in terms of $K_s$ and generates a signature state $|S\rangle=|\tilde{S}\rangle\otimes|\tilde{\Omega}\rangle$. Finally $|S\rangle$ and $|P\rangle$ are sent to the receiver Bob.
\item Bob implements the verification process by the following four steps: (1) Bob checks whether the state $|S\rangle$ is a $2k$-particle QECC by performing a syndrome measurement on it. (2) In terms of $K_v$, Bob decodes $|\tilde{S}\rangle$ as
    \begin{eqnarray}\label{eqa:1}
    K_v|S\rangle \rightarrow U|\tilde{S}\rangle    &=& J\|T\|^{1/2}\int\{|P\rangle_{r_1}|y_{r_{k+1}}\rangle_{r_2}|y_{r_{k+1}}\rangle_{r_{k+1}}...\nonumber\\
    & &{\  }\otimes|y_{r_{2k-1}}\rangle_{r_k}|y_{r_{2k-1}}\rangle_{r_{2k-1}}|\}dX\nonumber\\
    &=&
    J\|T\|^{1/2}|P\rangle_{r_1}|\Omega\rangle_{r_2,r_{k+1}}|\Omega\rangle_{r_3,r_{k+2}}...|\Omega\rangle_{r_k,r_{2k-1}}, \nonumber
    \end{eqnarray}
    where $J$ is the Jacobian for the transformation from $X$ to $y(X)$, and $|\Omega\rangle_{i,j}=\int_\mathbb{R}|y_l\rangle_i|y_l\rangle_j dx$ is an entanglement state of particles $i$ and $j$, $i=r_2,...,r_k, j,l=r_{k+1},...,r_{2k-1}$; (3) Bob verifies the entanglement properties of $k-1$ states $|\Omega\rangle_{r_2,r_{k+1}},|\Omega\rangle_{r_3,r_{k+2}},...,|\Omega\rangle_{r_k,r_{2k-1}}$, respectively. (4) Bob checks whether the decoded message state is the same as the received message state, and tests the equality of the decoded two-particle entangled state $|\Omega\rangle_{r_2,r_{k+1}}$ and the received two-particle entangled state $|\tilde{\Omega}\rangle$.
    If there is a failure in any step, Bob will reject $|S\rangle$ and stop the protocol.
\end{itemize}

\subsection{Cryptanalysis of the reviewed true QS scheme}
 The above scheme used for signing quantum messages \cite{Zeng:CVQS:IJQI07} does not involve a trustable arbitrator to help the receiver verify the signature.  Any receiver who is not always trustworthy can verify the validity of the signature directly. This scheme obviously violates Barnum \emph{et al}.'s conclusion \cite{Barnum:AQM:FOCS02} which stated that signing quantum messages is impossible since any scheme which allows one receiver to read a quantum message also allows the receiver to modify the message without the risk of being detected, and therefore all potential receivers of an authenticated message must be trustworthy. Therefore, if Barnum \emph{et al}.'s conclusion is right, Zeng \emph{et al}.'s scheme cannot be secure; or the other way round.

 In the following, we show that Barnum \emph{et al}.'s conclusion also adapts to Zeng \emph{et al}.'s scheme and the scheme is insecure. Firstly, we assume the receiver Bob gets the message $P$ and the corresponding valid signature $|S\rangle=|\tilde{S}\rangle\otimes|\tilde{\Omega}\rangle$ using Zeng \emph{et al}.'s scheme.  We then show Bob can forge a valid signature $|S'\rangle$ of another message $P'$ beneficial to him using the following three steps:
 \begin{itemize}
\item After decoding $|\tilde{S}\rangle$ using the way expressed in Eq. (\ref{eqa:1}), Bob replaces the decoded message state $|P\rangle$ with a new message state $|P'\rangle$, and the state of the whole system is changed to $|\Phi\rangle=J\|T\|^{1/2}|P'\rangle_{r_1}|\Omega\rangle_{r_2,r_{k+1}}|\Omega\rangle_{r_3,r_{k+2}}...|\Omega\rangle_{r_k,r_{2k-1}}$.
\item Bob applies $U^\dagger$, the inverse transformation of $U$, on $|\Phi\rangle$ to get $|\tilde{S}'\rangle=U^\dagger|\Phi\rangle$.
\item Bob generate the signature state $|S'\rangle=|\tilde{S}'\rangle\otimes|\tilde{\Omega}\rangle$ of $|P'\rangle$ by combing $|\tilde{S}'\rangle$ and
$|\tilde{\Omega}\rangle$.
\end{itemize}
The new message-signature pair $(P',|S'\rangle)$ is valid as it can be shown to pass the four steps of the verification phase due to
the following points: 1) $|S\rangle$ Bob holds is supposed to be a valid signature, so the entanglement properties of $|\Omega\rangle_{r_2,r_{k+1}},|\Omega\rangle_{r_3,r_{k+2}},...,|\Omega\rangle_{r_k,r_{2k-1}}$ are kept and the decoded state $|\Omega\rangle_{r_2,r_{k+1}}$ will be the same as $|\tilde{\Omega}\rangle$. That means Step (3) and Step (4) can be passed; 2) Due to $U|\tilde{S}'\rangle=UU^{\dagger}|\Phi\rangle=|\Phi\rangle$, Step (2) should also be passed; 3) Suppose $|S''\rangle=|\tilde{S}''\rangle\otimes|\tilde{\Omega}\rangle$ is the correct signature of $|P'\rangle$, then $|\tilde{S}''\rangle$ must be a $2k$-particle QECC. As $U|\tilde{S}''\rangle=|\Phi\rangle=U|\tilde{S}'\rangle$, $|\tilde{S}'\rangle$ is identical to $|\tilde{S}''\rangle$ and also is a $2k$-particle QECC which implies that Step (1) will be passed.

\section{Conclusion}

In this paper, we have shown arbitrated QS does not disobey Barnum \emph{et al}.'s conclusion about the impossibility of QS for quantum messages \cite{Barnum:AQM:FOCS02}, and the arbitrated QS under the proposed framework is possible to sign quantum messages with unconditional security. In addition, we have also explained that the existing true QS scheme presented by Zeng \emph{et al}. \cite{Zeng:CVQS:IJQI07} cannot get away from the restriction of Barnum \emph{et al}.'s conclusion since the scheme is shown to be insecure. Nevertheless, Barnum \emph{et al.}'s negative result does not preclude the possibility of QS for classical messages. So, how to construct efficient QS schemes to sign classical messages deserve further work.

\section*{Acknowledgement}

This work was partially supported by Natural Science Foundation of China (Grant Nos. 61202398, 61272295, and 61070232), Internal Research Grant of The Hong Kong Institute of Education (Grant No. RG 66/11-12), and the Foundation for Distinguished Young Talents in Higher Education of Guangdong (Grant No. LYM11093).

%\bibliographystyle{spphys}
%\nocite*
%\bibliography{AQSS}
\end{document}